\documentclass[12pt]{article}
\usepackage{amsmath}
\usepackage{longtable}
\usepackage{graphicx,psfrag,epsf}
\usepackage{enumerate}
\usepackage{natbib}
\usepackage{setspace}
\usepackage{hyperref} 
\usepackage{authblk}
\usepackage{hyperref}
\usepackage{booktabs}
\usepackage{amsmath,amssymb,amsthm}
\usepackage{graphicx,psfrag,epsf}
\usepackage{enumerate,booktabs,multirow,longtable,adjustbox,url}
\usepackage{algorithm,algpseudocode}
\usepackage{xr}
\usepackage{natbib}
\usepackage{hyperref}
\usepackage{subcaption}
\DeclareUnicodeCharacter{2212}{-}
\usepackage{url} 
\newcommand{\blind}{0}
\usepackage{amssymb}
\newcommand{\bfy}{\mathbf{y}}

\newcommand{\bfbeta}{\boldsymbol{\beta}}
\newcommand{\bfA}{\mathbf{A}}

\providecommand{\bfbeta}{\boldsymbol{\beta}}

\providecommand{\bfA}{\mathbf{A}}

\newcommand{\R}{\mathbb{R}}

\newcommand{\norm}[1]{\left\lVert#1\right\rVert}

\newcommand{\airbnbp}{Airbnb}

\makeatletter
\newcommand*{\addFileDependency}[1]{
  \typeout{(#1)}
  \@addtofilelist{#1}
  \IfFileExists{#1}{}{\typeout{No file #1.}}
}
\makeatother

\setcitestyle{authoryear,open={(},close={)}}
\date{} 
\begin{document}

\def\spacingset#1{\renewcommand{\baselinestretch}%
{#1}\small\normalsize} \spacingset{1}


\if0\blind
{
  \title{\bf Bayesian Shrinkage in High-Dimensional VAR Models: A Comparative Study}
\author[1,2]{Harrison Katz}
\author[3]{Robert E. Weiss}

\affil[1]{Department of Statistics, UCLA}
\affil[2]{Forecasting, Data Science, \airbnbp}

\affil[3]{Department of Biostatistics, UCLA Fielding School of Public Health }
  \maketitle
} \fi

\if1\blind
{
  \bigskip
  \bigskip
  \bigskip
  \begin{center}
    {\LARGE\bf Bayesian Shrinkage in High-Dimensional VAR Models: A Comparative Study}
\end{center}
  \medskip
} \fi

\begin{abstract}
High-dimensional vector autoregressive (VAR) models offer a versatile framework for multivariate time series analysis, yet face critical challenges from over-parameterization and uncertain lag order. In this paper, we systematically compare three Bayesian shrinkage priors (horseshoe, lasso, and normal) and two frequentist regularization approaches (ridge and nonparametric shrinkage) under three carefully crafted simulation scenarios. These scenarios encompass (i) overfitting in a low-dimensional setting, (ii) sparse high-dimensional processes, and (iii) a combined scenario where both large dimension and overfitting complicate inference.

We evaluate each method in quality of parameter estimation (root mean squared error, coverage, and interval length) and out-of-sample forecasting (one-step-ahead forecast RMSE). Our findings show that local-global Bayesian methods, particularly the horseshoe, dominate in maintaining accurate coverage and minimizing parameter error, even when the model is heavily over-parameterized. Frequentist ridge often yields competitive point forecasts but underestimates uncertainty, leading to sub-nominal coverage. A real-data application using macroeconomic variables from Canada illustrates how these methods perform in practice, reinforcing the advantages of local-global priors in stabilizing inference when dimension or lag order is inflated. 


\end{abstract}

\section{Introduction}

Vector autoregressive (VAR) models remain a cornerstone of multivariate time‑series analysis \citep{Sims1980, chan2020large, KOOP2013185}.  A $d$‑dimensional VAR($p$) posits that an observed series $\{\bfy_t\}_{t=1}^{T}$ satisfies
\begin{equation}
\label{eq:varp}
\bfy_t \;=\; \bfA_1\bfy_{t-1} + \bfA_2\bfy_{t-2} + \cdots + \bfA_p\bfy_{t-p} + \boldsymbol{\varepsilon}_t,
\end{equation}
where $\boldsymbol{\varepsilon}_t\!\sim\!\mathcal{N}(\mathbf{0},\Sigma_\varepsilon)$ is white noise.  
VARs are ubiquitous in macroeconomics and finance; see \citet{stock2002macroeconomic,CrumpEtAl2021_FRBNY,Carriero2022,ZhouChan2023} for recent surveys and methodological refinements.

A VAR becomes non‑stationary when the roots of $(I_d-\bfA_1z-\cdots-\bfA_p z^{p})$ lie on or inside the unit circle.  
Standard remedies include differencing, error‑correction representations, or Bayesian priors that down‑weight explosive parameter draws.

Because a VAR($p$) contains $d^{2}p$ coefficients, dimensionality rises quickly with either $d$ or $p$ \citep{Banbura2010,koop2013forecasting,korobilis2019adaptive}.  
Shrinkage—penalising or down‑weighting coefficients toward simpler structures—therefore plays a critical role in stabilising estimation and forecasts \citep{stock2002macroeconomic,chan2021minnesota}, especially when $d^{2}p \gg T$ \citep{bai2022macroeconomic}.

\medskip
\noindent\textbf{Bayesian shrinkage.}  
Local–global priors such as the horseshoe \citep{Carvalho2010} or Bayesian lasso \citep{Park2008} shrink most coefficients strongly while allowing a few to escape toward their likelihood values.  
Variants include the spike‑and‑slab \citep{george1997variable}, hierarchical horseshoes \citep{makalic2016simple,pruser2021horseshoe}, order‑invariant priors \citep{ChanKoopYu2024}, dynamic sparsification for hybrid TVP‑VARs \citep{Chan2023}, and structured factor‑augmented horseshoes \citep{ZhouChan2023}.  
Semi‑global or block‑specific shrinkage extends the idea by letting related groups share a common scale parameter \citep{GRUBER2025,PrueserBlagov2022}.  
Computational advances now make these priors feasible in very large systems: fast variational Bayes \citep{Bernardi2024}, approximate Bayes for huge multi‑country VARs \citep{Huber2022}, and conjugate subspace shrinkage \citep{HuberKoop2023}.  
Pandemic‑robust “outlier‑aware’’ priors improve real‑time performance during extreme events \citep{CascaldiGarcia2022}. Other recent Bayesian shrinkage methods in time series have been explored \citep{HuberKoop2023,kowal2019dynamic,KATZ20241556,forecast7030032}.

\medskip
\noindent\textbf{Frequentist and machine‑learning regularisation.}  
Ridge regression \citep{Doan1984} remains popular, while equation‑by‑equation Lasso or elastic‑net penalties \citep{Tibshirani1996,zou2005elasticnet,SanchezGarcia2022} yield sparse high‑dimensional VARs with frequentist guarantees under weak dependence \citep{Masini2022}.  
Envelope and reduced‑rank ideas provide additional dimension‑reduction tools \citep{SamadiHerath2023,CubaddaHecq2022}, and non‑linear factor compression has recently been explored \citep{Klieber2024}.  
Structured penalties—group Lasso, hierarchical lag shrinkage, SCAD, MCP—continue to improve point forecasts and interpretability \citep{NicholsonMattesonBien2017,Basu2019,Nicholson2020bigvar,SongBickel2019,ChenChen2021}.  
New inferential results ensure valid impulse‑response analysis even after sparse estimation \citep{Krampe2023}.

\medskip
\noindent\textbf{Forecasting and empirical performance.}  
Time‑varying parameter BVARs with horseshoe or normal‑gamma shrinkage adapt well to structural change \citep{BittoFruhwirthSchnatter2019,Feldkircher2024}.  
Factor‑adjusted network VARs (FNETS) separate common and idiosyncratic dynamics to enhance both forecasts and network interpretation \citep{Barigozzi2024}.  
Large‑scale empirical comparisons consistently find that local–global or semi‑global priors outperform purely global shrinkage and many frequentist alternatives in density and point forecasts \citep{HuberFeldkircher2019,Aprigliano2020,GefangKoopPoon2023}.

\medskip

In this paper, we compare five shrinkage approaches for high-dimensional VAR estimation: three Bayesian priors (horseshoe, lasso, and normal) and two frequentist estimators (ridge and nonparametric shrinkage). We evaluate each method’s handling of over-parameterization in both low- and high-dimensional settings, using root mean squared error, interval coverage, interval length, and RMSE to assess parameter recovery and forecast performance.

We find that local-global priors—particularly the horseshoe—strike a strong balance between parsimony and flexibility, delivering accurate estimation and consistent coverage even in heavily overfitted scenarios. While ridge regression frequently provides competitive point forecasts, it underestimates uncertainty when the parameter space grows large. Meanwhile, nonparametric shrinkage, though computationally efficient, suffers from undercoverage in complex models.

Application to Canadian macroeconomic time series illustrates how each method behaves with various lag choices and a relatively small sample. These empirical findings corroborate the simulation evidence: local-global priors, especially the horseshoe, remain robust even when the chosen lag order exceeds what is strictly necessary.

The remainder of the paper is structured as follows. 
Section~2 defines the VAR($p$) framework and outlines both Bayesian and frequentist shrinkage estimators. 
Section~3 describes our simulation designs, metrics, and results. Section~4 provides the real-data application to Canadian macroeconomic variables.  Finally, we offer concluding remarks.

\section{Bayesian and Frequentist Approaches to VAR(\boldmath$p$)}
\label{sec:methodology}

We study a $d$-dimensional VAR($p$) of the form
\begin{equation}
\mathbf{y}_t \;=\; \mathbf{A}_1 \,\mathbf{y}_{t-1} \;+\; \cdots \;+\; \mathbf{A}_p \,\mathbf{y}_{t-p} \;+\; \boldsymbol{\varepsilon}_t,
\end{equation}
where each $\mathbf{A}_i$ is a $d \times d$ coefficient matrix, and $\boldsymbol{\varepsilon}_t$ is a white-noise process following
\[
\boldsymbol{\varepsilon}_t \;\sim\; \mathcal{N}\!\bigl(\mathbf{0}, \,\Sigma_\varepsilon\bigr),
\quad
\mathrm{Cov}\bigl(\boldsymbol{\varepsilon}_t, \boldsymbol{\varepsilon}_s\bigr) \;=\; \mathbf{0}
\text{ for } t \neq s.
\]
We treat each $\mathbf{y}_t$ as a $d\times1$ column vector.

\medskip

\noindent
\textbf{Vectorizing the Coefficients.}\;
Let 
\[
\mathbf{B} \;=\; \bigl[\mathbf{A}_1\;\mathbf{A}_2\;\cdots\;\mathbf{A}_p\bigr]
\;\in\;\mathbb{R}^{\,d\times(d\,p)},
\]
i.e., the horizontal concatenation of the $p$ coefficient matrices with elements $\beta_j$, $j=1, \dots, d \times dp$.  Then define the \emph{lagged-regressor} vector
\[
\mathbf{X}_t 
\;=\;
\begin{pmatrix}
\mathbf{y}_{t-1} \\
\mathbf{y}_{t-2} \\
\vdots\\
\mathbf{y}_{t-p}
\end{pmatrix}
\;\in\;\mathbb{R}^{\,d\,p\times 1},
\]
so that
\[
\mathbf{B}\,\mathbf{X}_t 
\;\in\;\mathbb{R}^{\,d\times1}.
\]
The VAR($p$) model in \eqref{eq:varp} can thus be written as
\[
\mathbf{y}_t
\;=\;
\mathbf{B}\,\mathbf{X}_t 
\;+\;
\boldsymbol{\varepsilon}_t,
\quad
\boldsymbol{\varepsilon}_t \sim \mathcal{N}\!\bigl(\mathbf{0}, \,\Sigma_\varepsilon\bigr).
\]

\subsection{Bayesian Shrinkage Priors}

In a fully Bayesian treatment, we specify priors for both the coefficient matrix and the error covariance. We gather the coefficients into a matrix 
\(
\mathbf{B}\in\mathbb{R}^{d \times (d\,p)},
\)
so that 
\[
\mathbf{y}_t 
\;\sim\; 
\mathcal{N}\!\bigl(\mathbf{B} \mathbf{X}_t^\prime ,\;\Sigma_\varepsilon\bigr),
\]
where $\mathbf{X}_t$ is the row vector of the lagged responses at time~$t$. To ensure $\Sigma_\varepsilon$ is positive-definite, we use a Cholesky-factor parameterization:
\[
\Sigma_\varepsilon
\;=\;
\mathbf{L}\,\mathbf{L}^\top,
\quad
\mathbf{L} = \text{diag}(\sigma) \,\mathbf{L}_\Omega,
\] where $\mathbf{L}_\Omega$ is the Cholesky factor of a correlation matrix with an LKJ prior \citep{Lewandowski2009}, and each component of $\sigma$ follows a half-Cauchy prior. This flexible structure permits correlation among the $d$ error components.

We then place shrinkage priors on each coefficient in~$\mathbf{B}$. Below, we detail three such priors---normal (ridge), horseshoe, and Bayesian lasso---all of which can be combined with the same LKJ-based prior for $\Sigma_\varepsilon$:

\paragraph{Normal Prior (Bayesian Ridge).}
A normal (Gaussian) prior imposes a global $\ell_2$ penalty. For each coefficient $\beta_j$ we set priors, 
\[
\beta_j \;\sim\; \mathcal{N}\!\bigl(0,\,1\bigr),
\quad
j=1,\ldots,d^2 p,
\]
so that most coefficients are moderately shrunk towards zero. This parallels the frequentist ridge penalty, and one can include an additional scale factor if stronger or weaker global shrinkage is desired depending on the data set.

\paragraph{Horseshoe Prior.}
The horseshoe prior \citep{Carvalho2010, makalic2016simple} introduces more adaptive shrinkage via a local-global hierarchy. Each $\beta_j$ is modeled as 
\(\beta_j = B_{\mathrm{raw},j}\,\lambda_j\,\tau\),
where 
\(B_{\mathrm{raw},j} \sim \mathcal{N}(0,1)\),
\(\lambda_j \sim \mathrm{C}^+(0,1)\) (local scale), and 
\(\tau \sim \mathrm{C}^+(0,1)\) (global scale). Small coefficients are heavily shrunk by small local scales, while the heavy-tailed Cauchy priors allow some large signals to remain.

\paragraph{Bayesian Lasso Prior.}
Finally, the Bayesian lasso \citep{Park2008} imposes a Laplace (double-exponential) prior,
\[
\beta_j  | \eta \;\sim\; \mathrm{Laplace}(0,\eta),
\]
which corresponds to an $\ell_1$ penalty in a frequentist setting. As with the horseshoe, this encourages coefficients to be near zero, possibly leading to sparsity in~$\bfbeta$.

In all three cases, the covariance matrix $\Sigma_\varepsilon$ is handled by the same LKJ-based prior, thus capturing potential correlations in the innovation process. The posterior distribution factors as
\[
\pi\bigl(\bf{B}, \Sigma_\varepsilon \mid \mathbf{y_t}\bigr)
\;\propto\;
\ell\bigl(\mathbf{y_t} \mid \bf{B}, \Sigma_\varepsilon\bigr)\,\pi(\bf{B})\,\pi(\Sigma_\varepsilon),
\]
where $\ell$ is the Gaussian likelihood induced by the VAR model, and $\pi(\bf{B})$, $\pi(\Sigma_\varepsilon)$ encode the chosen shrinkage and covariance priors, respectively. By jointly estimating $\mathbf{B}$ and $\Sigma_\varepsilon$, this framework avoids the assumption of uncorrelated errors and allows us to examine how different shrinkage priors influence coefficient estimation in a fully multivariate setting.

\paragraph{How the priors influence estimation.}
The three priors differ in the \emph{shape} and \emph{hierarchy} of their scale parameters, and
these choices translate directly into the amount and selectivity of shrinkage:

\begin{itemize}
  \item \textbf{Normal (ridge).} A single global variance forces \emph{equal} shrinkage on every
        $\beta_j$, pulling all coefficients toward~0 irrespective of their signal strength.
        This produces stable point estimates but can underestimate posterior uncertainty when
        $d^2p \gg T$.
  \item \textbf{Horseshoe.}  Independent local scales $\lambda_j\!\sim\!\mathrm{C}^+$ give each
        coefficient its \emph{own} amount of shrinkage, while the heavy‑tailed Cauchy hierarchy
        preserves large signals.  The result is strong suppression of noise coefficients together
        with wider credible intervals for the few active ones.
  \item \textbf{Bayesian lasso.}  The exponential tails of the Laplace prior lie between the two
        extremes above, producing more sparsity than the Normal but heavier global shrinkage
        than the Horseshoe; moderate signals are therefore attenuated the most.
\end{itemize}

These qualitative differences predict the empirical patterns we later observe: ridge delivers the
narrowest—but sometimes under‑covering—intervals, the horseshoe attains the best balance of low
RMSE and near‑nominal coverage, and the lasso sits in between.  Section~\ref{sec:discussion}
returns to this point in light of the simulation and data results.

\subsection{Frequentist Methods}

\paragraph{Ridge Regression.} 
Classical ridge regression for a VAR($p$) solves
\begin{equation}
\label{eq:ridge}
\min_{\bfbeta \in \R^{d^2p}}
\;\sum_{t=p+1}^T 
\norm{\bfy_t - \sum_{i=1}^p \bfA_i\,\bfy_{t-i}}^2
\;+\;
\lambda
\norm{\bf{B}}_{2}^2,
\end{equation}
where $\bf{B}$ is just the vectorized collection of $\{\bfA_i\}$. We set the regularization parameter $\lambda=.1$ in our analysis.

\paragraph{Nonparametric Shrinkage (NS).} We use a James--Stein-like shrinkage approach for VAR coefficients \citep{Giannone2015,DelNegro2015}, implemented in \verb|R| via \verb|VARshrink| with \texttt{method="ns"}. Instead of explicitly solving \eqref{eq:ridge}, the NS method estimates the necessary sample covariances of ${\bfy_{t-i},\bfy_t}$ and then applies a closed-form shrinkage rule to these covariance estimates, thereby producing a shrunk solution for $\bf{B}$. In this paper, we rely on the default choice for the shrinkage parameter, which \verb|VARshrink| selects via a moment-based (empirical Bayes) formula akin to Stein’s unbiased risk estimate.

\section{Simulation Studies}

\subsection{Data-Generating Processes}
\label{sec:dgp}

We examine three Monte--Carlo scenarios that differ in dimension~($d$) and in whether the
fitted lag order~$p$ coincides with the true lag order~$p^{\star}$.  
Table~\ref{tab:dgp_summary} records the fixed design choices; the recipe that follows
is applied independently in each of the $N_{\text{rep}}=50$ replications.

\begin{table}[htbp]
\centering\small
\caption{Key parameters of each simulation scenario.  
``Sparsity'' is the probability that an entry of~$A_{1}$ is set to~$0$.  
All series use a 50-observation burn-in, followed by
$T_{\mathrm{train}}=180$ training points and a $H=20$-point test set.}
\label{tab:dgp_summary}
\begin{tabular}{lccccc}
\toprule
Scenario & $d$ & true $p^{\star}$ & fitted $p$ & sparsity & $\sigma_\varepsilon^{2}$  \\
\midrule
1: low-$d$, over-fit      &  3 & 1 & 4 & 0.70 & 0.05 \\
2: high-$d$, well-fit     & 20 & 1 & 1 & 0.70 & 0.10 \\
3: high-$d$, over-fit     & 20 & 1 & 4 & 0.70 & 0.10 \\
\bottomrule
\end{tabular}
\end{table}

\begin{algorithm}[htbp]
\caption{Simulation procedure for a single replication} 
\label{alg:simulate}
\begin{algorithmic}[1]
  \Statex \textbf{Input:} $(d,p^{\star},\text{sparsity},\sigma_\varepsilon^{2})$
  \Statex \textbf{Output:} training sample $\{\mathbf y_t\}_{t=1}^{T_{\text{train}}}$ and
          hold-out sample $\{\mathbf y_t\}_{t=T_{\text{train}}+1}^{T_{\text{train}}+H}$
  \vspace{2pt}
  \State \textbf{Draw the coefficient matrix $A_{1}$.}
        Each entry $(A_{1})_{ij}$ is independently
        \[
          (A_{1})_{ij} \;=\;
          \begin{cases}
            0,                & \text{with probability } \text{sparsity},\\[4pt]
            U(-0.4,0.4),      & \text{otherwise.}
          \end{cases}
        \]
        All higher-order true lags are set to zero:
        $A_{2}=A_{3}=\dots=A_{p^{\star}}=\mathbf 0$.
  \smallskip
  \State \textbf{Stationarity margin.}
        Compute the spectral radius $\rho_{\max}$ of $A_{1}$ and rescale
        $A_{1}\leftarrow A_{1}/(1.1\,\rho_{\max})$ so that all roots of
        $I_d-A_{1}z$ are strictly inside the unit circle.
  \smallskip
  \State \textbf{Simulate the innovations.}
        Draw $\boldsymbol\varepsilon_t\sim\mathcal N(\mathbf 0,\sigma_\varepsilon^{2}I_d)$
        independently for $t=-49,\dots,T_{\text{train}}+H$.
  \smallskip
  \State \textbf{Generate the series.}
        Starting from $\mathbf y_{-50}=\mathbf 0$, evolve
        $\displaystyle \mathbf y_t = A_{1}\mathbf y_{t-1} + \boldsymbol\varepsilon_t$.
  \smallskip
  \State \textbf{Burn-in.}
        Discard the first 50 observations ($t=-49,\dots,0$); the retained
        series therefore begins at $t=1$.
  \smallskip
  \State \textbf{Split the data.}
        Keep $t=1,\dots,T_{\text{train}}$ for estimation and
        $t=T_{\text{train}}+1,\dots,T_{\text{train}}+H$ for forecast evaluation.
\end{algorithmic}
\end{algorithm}

\paragraph{Stationarity margin.}
Following the textbook discussion in \citet[][Sec.~2.1]{Lutkepohl2005},
we scale the simulated companion matrix so that its spectral radius is
comfortably inside the unit circle.  Specifically, we divide the
initial draw $A_{1}$ by $1.1\,\rho_{\max}$, where $\rho_{\max}$ is the
largest eigenvalue (in modulus).  This simple rescaling keeps all roots
of the characteristic polynomial strictly below one—thereby ensuring
covariance–stationarity—while preserving empirically plausible
coefficient magnitudes for macro‑VAR applications.

\subsection{Estimation Methods}
\paragraph{Design Matrix Setup.}
To estimate a VAR($p$) in a linear regression framework, we arrange lagged responses into a design matrix 
\(\mathbf{X}\in\mathbb{R}^{(T_{\mathrm{train}} - p)\times(d\,p)}\). 
For \(t = p+1,\dots,T_{\mathrm{train}}\), the \(t\)-th row of \(\mathbf{X}\) (denoted \(\mathbf{X}_t^\top\)) is formed by horizontally concatenating the transposes of the \(p\) lagged column vectors 
\[
  \mathbf{y}_{t-1}, \mathbf{y}_{t-2}, \ldots, \mathbf{y}_{t-p}
  \quad
  (\text{each } d\times1).
\]
Hence, each row of \(\mathbf{X}\) is a \(1\times(d\,p)\) vector. 
Likewise, the \(t\)-th row of the response matrix 
\(\mathbf{Y}\in\mathbb{R}^{(T_{\mathrm{train}} - p)\times d}\)
is simply the transpose \(\mathbf{y}_t^\top\in\mathbb{R}^{1\times d}\).
Once \(\mathbf{X}\) and \(\mathbf{Y}\) are formed, any penalized or Bayesian regression method can be applied directly, 
and the estimated coefficient matrix is then reshaped to match 
\(\mathbf{B}\in\mathbb{R}^{d\times(d\,p)}\)
as defined in Section~\ref{sec:methodology}.

\paragraph{Frequentist fits and block-bootstrapped standard errors.}
We estimate the VAR coefficients in Ridge (\texttt{glmnet} with \(\alpha=0\)) and NS (\texttt{VARshrink} with \texttt{method="ns"}) by penalized least squares, and then obtain empirical standard errors via a \emph{block bootstrap} to better respect local time dependence. Specifically we

\begin{enumerate}
  \item \textbf{Partition into blocks:} We group the training data \(\{X, Y\}\) into non-overlapping consecutive blocks of length 4. 
  We choose a block size of 4 because our VAR models use up to 4 lags, so each block captures the short-range autocorrelation structure of interest. 
  \item \textbf{Sample blocks with replacement:} To form a bootstrap dataset of the same size as the original, we randomly select blocks \emph{with replacement} until we have at least \(T_{\mathrm{train}}\) observations in total. 
  \item \textbf{Refit the model:} We refit the Ridge or NS model on this resampled dataset and record the estimated coefficients.
  \item \textbf{Repeat and aggregate:} Steps (2)--(3) are repeated for 30 bootstrap replications (\(\texttt{n\_boot = 30}\)). The empirical standard error for each coefficient is then taken to be the sample standard deviation of its estimates across these replications.
\end{enumerate}

This procedure retains within-block autocorrelations (up to 4 lags) while randomly mixing which blocks are selected, preserving important time-series structure better than naive row-wise (i.i.d.) resampling. As a result, the resulting intervals yield more realistic coverage for dependent data.

\paragraph{Bayesian fits.}
We fit the three Bayesian models by calling \texttt{Stan} with 4 parallel Markov chains, each run for 2000 total iterations (the first 500 of which are warm-up). We fix \texttt{seed=123} for reproducibility and use $\{\texttt{adapt\_delta}=0.9, \;\texttt{max\_treedepth}=12\}.$ In Stan’s Hamiltonian Monte Carlo (HMC) framework, \texttt{adapt\_delta} is the target acceptance probability, and increasing it to 0.9 aims for smaller step sizes and more conservative sampling. The \texttt{max\_treedepth} parameter caps the depth of the binary tree in each iteration’s leapfrog integrator, preventing extremely long trajectories.

For each chain, we obtain posterior draws of the coefficient vector \(\bfbeta\). We summarize each coefficient by its posterior mean and 95\% central credible interval (2.5\% and 97.5\% quantiles).

\subsection{Performance Metrics}

We evaluate each method along two dimensions: \emph{parameter estimation} and \emph{forecast} performance.

\paragraph{Parameter estimation.}
Let \(\bfbeta_{\mathrm{true}}\) denote the true parameters in $\bf{B}$. Each frequentist method (Ridge or Nonparametric Shrinkage) estimates \(\bfbeta\) by minimizing a penalized least squares criterion, whereas each Bayesian method (Normal/Ridge, Lasso, Horseshoe) uses the posterior mean from MCMC samples as \(\widehat{\bfbeta}\). We then compute the root mean squared error (RMSE),
\[
\mathrm{RMSE} \;=\;
\sqrt{\frac{1}{d^2 p}
\sum_{j=1}^{d^2 p} 
\bigl(\widehat{\beta}_j - \beta_{j,\mathrm{true}}\bigr)^2},
\]
to measure how closely \(\widehat{\bfbeta}\) matches \(\bfbeta_{\mathrm{true}}\).

Next, we construct 95\% intervals for each coefficient by applying a block bootstrap to estimate standard errors and forming approximate normal intervals of the form 
\(\widehat{\beta}_j \pm z_{0.975}\,\mathrm{SE}_j\) for the frequentist approaches, whereas for the Bayesian methods we use the 2.5\% and 97.5\% posterior quantiles from the MCMC samples. We record the \emph{empirical coverage} (the percentage of intervals that contain the true value) and the \emph{average interval length} to assess how well each approach quantifies uncertainty.

\paragraph{Forecasting performance.}
To assess predictive accuracy, we reserve the final 20 observations as a test set. Each method then produces sequential one-step-ahead forecasts by estimating \(\mathbf{y}_{t+1}\) at time \(t\) based on all data up to \(\mathbf{y}_{t}\), avoiding the accumulation of multi-step errors. We compute the average root mean squared forecast error (RMSE)
\[
\mathrm{Forecast\,RMSE} 
\;=\;
\sqrt{
\frac{1}{20\,d}
\sum_{t=T_{\mathrm{train}}+1}^{T_{\mathrm{train}}+20}
\norm{\mathbf{y}_t - \hat{\mathbf{y}}_t}_2^2
},
\]
where \(\hat{\mathbf{y}}_t\) is the forecast at time \(t\). After 50 replications per scenario, we summarize the average forecasting RMSE and coverage to compare each method’s predictive capabilities.

\subsection{Simulation Results}
\label{sec:simresults}

\begin{figure}[ht!]
\centering
\includegraphics[width=\textwidth]{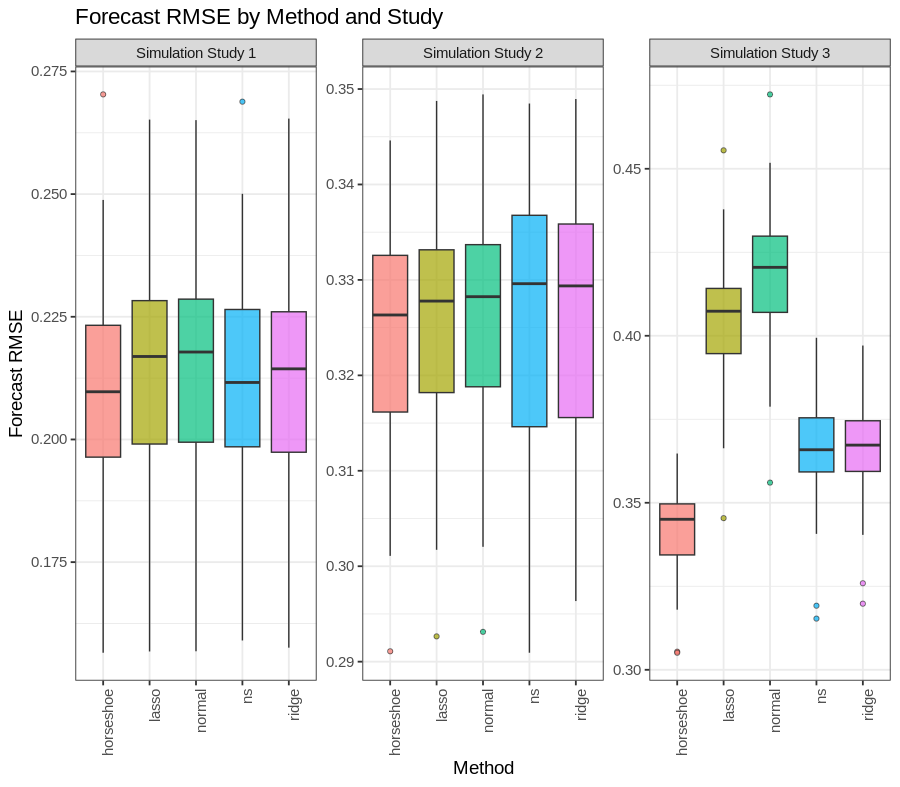}
\caption{%
\textbf{Forecast RMSE by Method and Study (All Coefficients).} Boxplots reflect the distribution of one-step-ahead RMSE across the 50 replications. Horseshoe achieves or ties for the lowest forecast error, especially in the high-dimension overfit scenario (Study~3).%
}
\label{fig:forecast_rmse_plot}
\end{figure}

\begin{figure}[ht!]
\centering
\includegraphics[width=\textwidth]{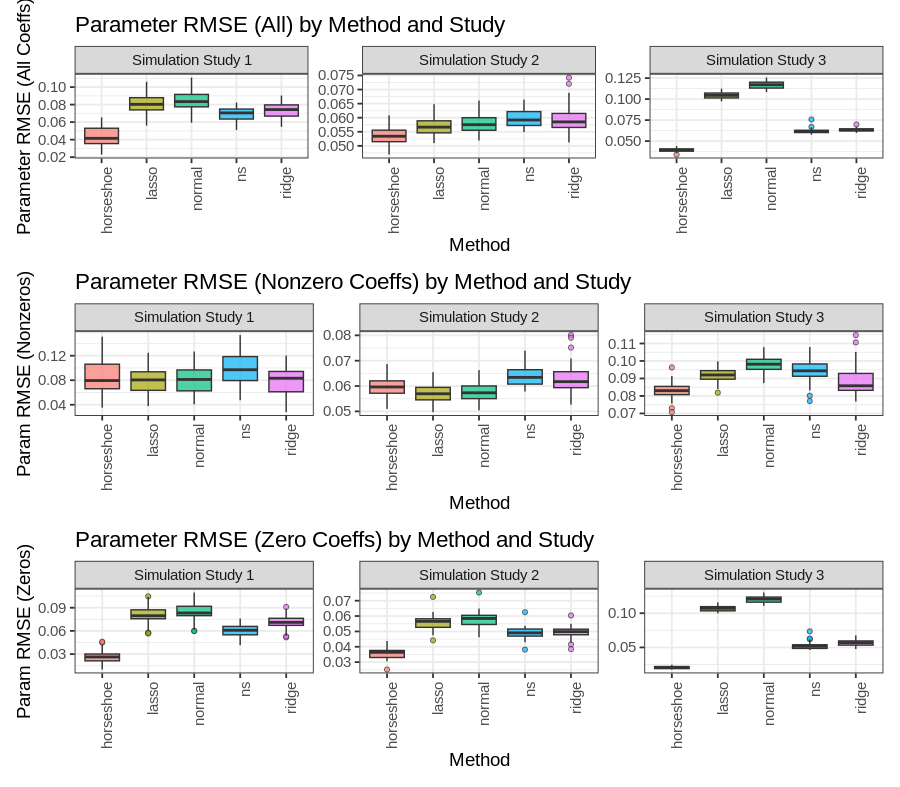}
\caption{%
\textbf{Parameter RMSE by Method and Study.}
Horseshoe is consistently lowest in overall parameter RMSE, while NonparamShrink (\emph{ns}) occasionally performs well but can exhibit greater variance or undercoverage.%
}
\label{fig:param_rmse_plot}
\end{figure}

\begin{figure}[ht!]
\centering
\includegraphics[width=\textwidth]{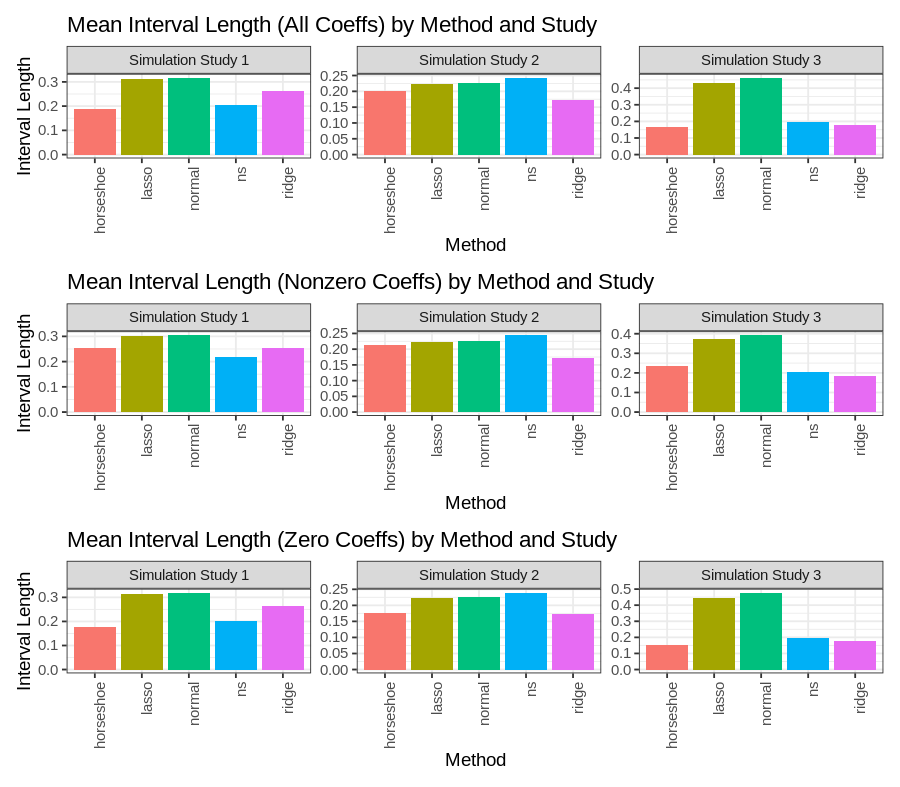}
\caption{%
\textbf{Mean Interval Length  by Method and Study.}
Shorter intervals may indicate overconfidence if coverage is below the nominal 95\%; for instance, \emph{ns} has narrower intervals but lower coverage in some scenarios.%
}
\label{fig:interval_length_plot}
\end{figure}

\begin{figure}[ht!]
\centering
\includegraphics[width=\textwidth]{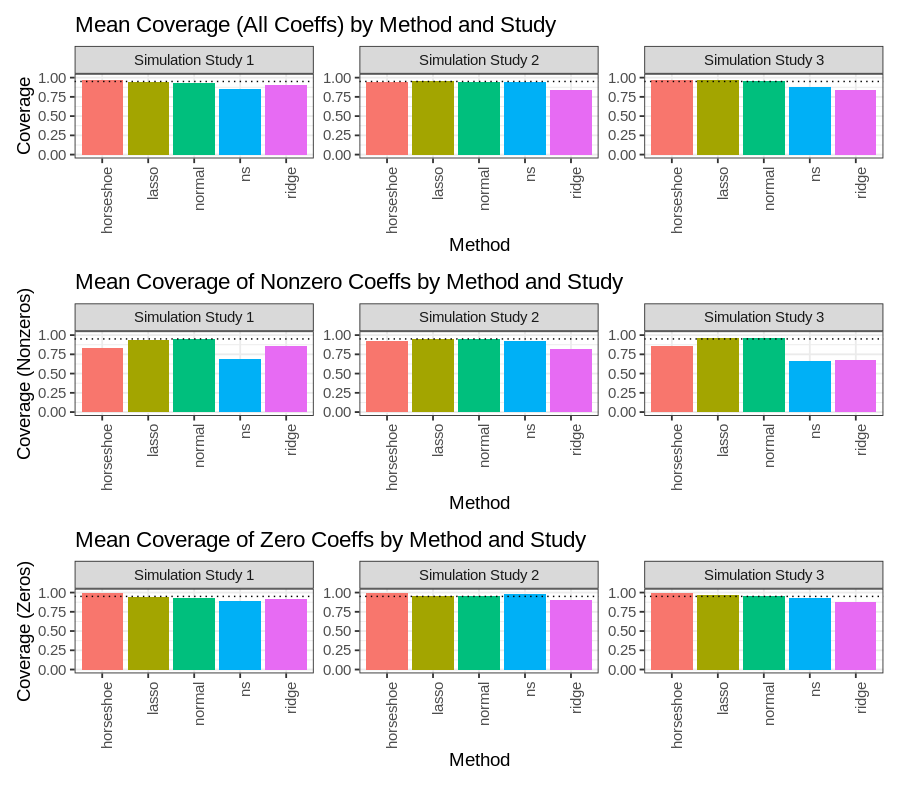}
\caption{%
\textbf{Coverage by Method and Study.}
A dotted line at $0.95$ indicates the nominal coverage target. Horseshoe, Lasso, and Normal usually achieve near 95\%, while \emph{ns} and Ridge can dip below this level for high-dimensional or overfit scenarios.%
}
\label{fig:coverage_plot}
\end{figure}

The results from the three simulation studies are shown in tables~\ref{tab:summary_all}--\ref{tab:summary_nonzero} and figures~\ref{fig:forecast_rmse_plot}--\ref{fig:coverage_plot}.  The percentage of replications in which each method has the lowest forecast RMSE or parameter RMSE is shown in table~\ref{tab:times_best}.

\subsubsection{Scenario 1 (Low-Dimension, Overfit Lag)}

\paragraph{Forecasting}
The top block of Table~\ref{tab:summary_all} has each method’s mean forecast RMSE. Horseshoe has the smallest value (0.211), followed by \emph{ns} and Ridge (0.213), Lasso (0.214), and Normal (0.215). Table~\ref{tab:times_best} has the proportion of replications in which each method has the best forecast: Horseshoe leads with 60\%, \emph{ns} has 20\%, Normal 10\%, Ridge 6\%, and Lasso 4\%.

\paragraph{Parameter Estimation}
Horseshoe has the lowest overall parameter RMSE (0.0434) and exceeds average nominal coverage (97.2\%), with intervals about 8\% shorter than those of the next-best method. Lasso (0.0803) and Normal (0.0838) have higher RMSEs but maintain coverage near 94--95\%. Both \emph{ns} (0.0693) and Ridge (0.0730) occupy a middle tier; \emph{ns} has coverage of 85.7\% but yields narrower intervals (mean length 0.204). Horseshoe has the best parameter RMSE in all replications (100\%), giving strong shrinkage without sacrificing coverage.

\subsubsection{Scenario 2 (High-Dimension, Correct Lag)}

\paragraph{Forecasting}
All methods have similar forecasting accuracy (middle block of Table~\ref{tab:summary_all}). Horseshoe has the smallest mean forecast RMSE (0.325), followed by Lasso (0.326) and Normal, \emph{ns}, and Ridge (0.327). Horseshoe is the top forecaster in 48\% of replications, Lasso and \emph{ns} each in 20\%, Ridge in 10\%, and Normal in 2\% (Table~\ref{tab:times_best}).

\subsubsection*{Parameter Estimation}
Horseshoe again has the lowest parameter RMSE (0.0536). Lasso, Normal, \emph{ns}, and Ridge cluster between 0.0568 and 0.0598. Coverage remains high (94--95\%) for Horseshoe, Lasso, Normal, and \emph{ns}, but dips to 84\% for Ridge. Table~\ref{tab:summary_zero} has results for zero coefficients, where Horseshoe has an RMSE of 0.0357 and 99.0\% coverage. The \emph{ns} method handles zero parameters well but sometimes underperforms on nonzeros. Horseshoe has a nonzero RMSE of 0.0596, higher than Lasso and Normal (0.0571--0.0576), while maintaining overall coverage of 92.9\%. Horseshoe has the best parameter RMSE in 90\% of replications, followed by Ridge in 10\% (Table~\ref{tab:times_best}).

\subsubsection{Scenario 3 (High-Dimension, Overfit Lag)}

\paragraph*{Forecasting}
In the bottom block of Table~\ref{tab:summary_all}, Horseshoe has the smallest mean forecast RMSE (0.342), followed by \emph{ns} and Ridge (0.365--0.366), Lasso (0.404), and Normal (0.418). Horseshoe is the top forecaster in 100\% of replications (Table~\ref{tab:times_best}), indicating a strong ability to handle overfitting.

\paragraph*{Parameter Estimation}
Horseshoe has the lowest parameter RMSE (0.0394), with better-than-nominal average coverage (97.5\%) and intervals about 8\% shorter than those of the next-best method. Lasso (0.104) and Normal (0.117) have higher RMSEs but maintain nominal coverage (95--96\%) through wider intervals (0.432--0.464). Both \emph{ns} and Ridge have moderate RMSEs (0.0619--0.0635) but show lower coverage (88.2\% and 84.5\%) and narrower intervals (0.182--0.197). Horseshoe remains the top performer in parameter RMSE for 100\% of the replications.

Overall, Horseshoe consistently has excellent forecast accuracy and parameter recovery, including the lowest RMSE, high coverage, and moderate interval lengths. Ridge occasionally has strong forecasts but frequently undercovers in higher dimensions. Lasso and Normal have intermediate performance for both forecasting and parameter estimation, ensuring reasonable coverage by using somewhat larger intervals. The \emph{ns} approach is computationally efficient and sometimes has precise point estimates, but coverage can be volatile due to overly narrow intervals. These findings reinforce the advantages of local-global shrinkage (Horseshoe) in moderate- and high-dimensional VAR contexts, especially when the lag order is inflated.

\section{Data Analysis}
\label{sec:data}

We illustrate our methods on the \texttt{Canada} dataset from the \textsf{R} package \textbf{vars} \citep{JSSv027i04}, which provides quarterly macroeconomic observations on four Canadian variables spanning $T=84$ quarters (1980Q1--2000Q4): employment (\emph{e}, in log-index form), productivity (\emph{prod}, in log-index form measuring labor productivity), real wages (\emph{rw}, in log-index form), and the unemployment rate (\emph{U}, in percent). Economic considerations suggest these variables are jointly dependent, making a Vector Auto Regression (VAR)-based approach suitable.

\paragraph{Differencing and Stationarity.}
To reduce nonstationarity, we difference each series once
\[
\Delta \mathbf{y}_{t} \;=\; \mathbf{y}_{t} \;-\; \mathbf{y}_{t-1}
\quad (t=2,\dots,T).
\]
We then estimate the VAR on these differenced observations. To obtain forecasts on the original scale, we \emph{invert} the differencing by recursively summing the predicted differences
\[
\widehat{\mathbf{y}}_{T+1}
\;=\;
\mathbf{y}_{T}
\;+\;
\widehat{\Delta \mathbf{y}}_{T+1},
\quad
\widehat{\mathbf{y}}_{T+2}
\;=\;
\widehat{\mathbf{y}}_{T+1}
\;+\;
\widehat{\Delta \mathbf{y}}_{T+2},
\quad
\text{etc.}
\]
This ensures the final forecasts reflect the original scale of the data, while the differencing step helps achieve stationarity and limit spurious trends when fitting the VAR.

\paragraph{Lag–order experiment ($p=1,\dots,12$).}
For each shrinkage method we fit twelve separate VAR($p$) models
($p=1,\dots,12$).  Every model is trained on the first $T-4$ quarterly
differences and produces one‑step‑ahead level forecasts for 2000Q1–Q4.
For a given lag order $p$ we compute
\[
  \mathrm{RMSE}_{p}
  \;=\;
  \sqrt{\frac{1}{4}\sum_{h=1}^{4}\bigl(y_{T+h}-\hat y^{(p)}_{T+h}\bigr)^{2}},
  \qquad
  \mathrm{MAPE}_{p}
  \;=\;
  \frac{100}{4}\sum_{h=1}^{4}
  \Bigl|\frac{y_{T+h}-\hat y^{(p)}_{T+h}}{y_{T+h}}\Bigr|.
\]
Hence each method yields a \emph{set} of twelve numbers
$\{\mathrm{RMSE}_{p}\}_{p=1}^{12}$ and
$\{\mathrm{MAPE}_{p}\}_{p=1}^{12}$.
We summarize these sets by their \emph{means}
\[
  \overline{\mathrm{RMSE}}
    = \frac{1}{12}\sum_{p=1}^{12}\mathrm{RMSE}_{p},
  \quad
  \overline{\mathrm{MAPE}}
    = \frac{1}{12}\sum_{p=1}^{12}\mathrm{MAPE}_{p},
\]
and by their sample standard deviations
\[
  \mathrm{SD\,RMSE}
    = \sqrt{\frac{1}{11}\sum_{p=1}^{12}
              \bigl(\mathrm{RMSE}_{p}-\overline{\mathrm{RMSE}}\bigr)^{2}},
  \quad
  \mathrm{SD\,MAPE}
    = \sqrt{\frac{1}{11}\sum_{p=1}^{12}
              \bigl(\mathrm{MAPE}_{p}-\overline{\mathrm{MAPE}}\bigr)^{2}}.
\]
These four statistics: $\overline{\mathrm{RMSE}}$, $\overline{\mathrm{MAPE}}$,
$\mathrm{SD\,RMSE}$ and $\mathrm{SD\,MAPE}$ provide, respectively, the
\emph{average forecast accuracy} and the \emph{across‑lag variability} reported
in Table~\ref{tab:canada_summary_p1to10}.

\subsection{Results}
\subsubsection{Aggregate Performance}
Overall, \emph{Horseshoe} achieves the most consistent and accurate forecasts, attaining the lowest mean $  \overline{\mathrm{RMSE}}$ (0.51). \emph{Ridge} ($  \overline{\mathrm{RMSE}}$ = 0.56) and \emph{NS} ($  \overline{\mathrm{RMSE}}$ = 0.60) provide somewhat intermediate performance, while \emph{Lasso} (RMSE = 0.60) and \emph{Normal} ($  \overline{\mathrm{RMSE}}$ = 0.63) tend to yield slightly larger prediction errors. In terms of variability, Horseshoe’s standard deviation of RMSE (0.22) is comparable to that of Lasso and Ridge, whereas Normal exhibits the largest SD (0.26).

\begin{figure}[ht!]
    \centering
    \includegraphics[width=1.1\textwidth]{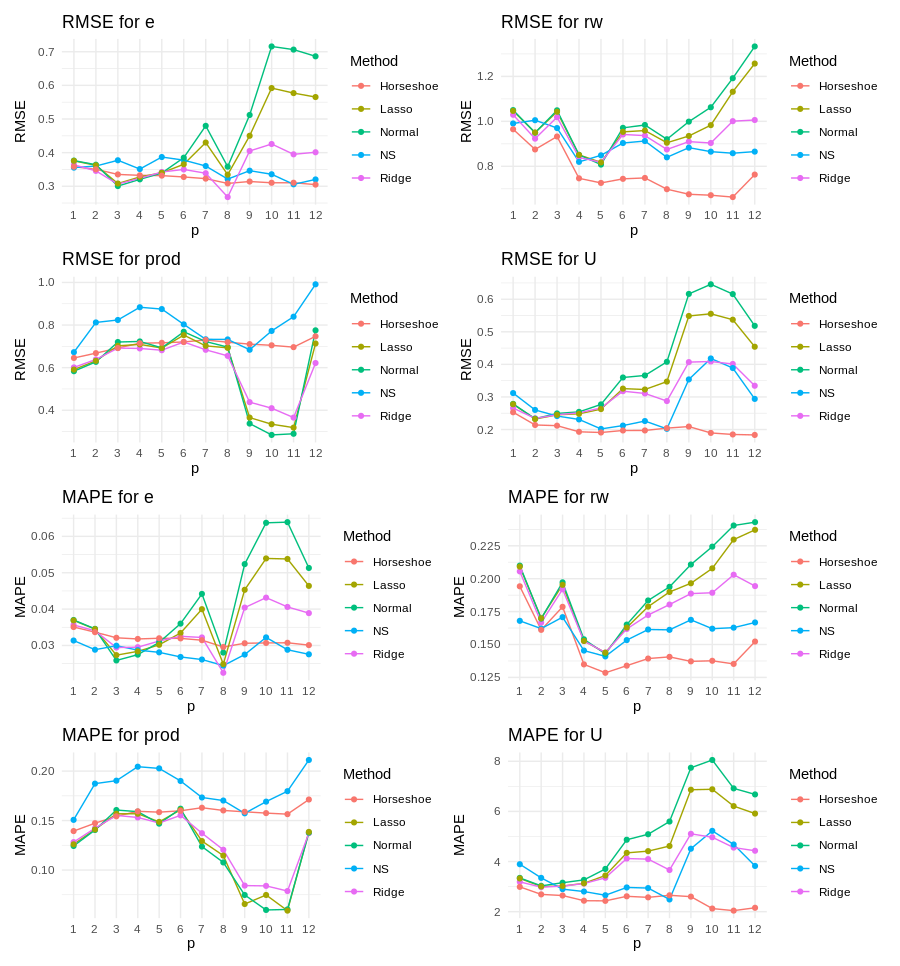}
    \caption{Out-of-sample forecasting accuracy for each Canadian macroeconomic variable—employment ($e$), real wages ($rw$), productivity ($prod$), and unemployment ($U$)—across increasing VAR orders ($p=1,\dots,12$). 
    The top row shows the Root Mean Squared Error (RMSE), and the bottom row shows the Mean Absolute Percentage Error (MAPE). 
    Each colored line corresponds to one of five shrinkage methods (Horseshoe, Lasso, Normal, ns, and Ridge). 
    Overall, Horseshoe achieves the lowest MAPE and exhibits relatively stable performance as $p$ increases. 
    Lasso and Ridge  perform well but show greater variability at higher orders.}
    \label{fig:canada_rmses_mapes_series}
\end{figure}

For $ \overline{\mathrm{MAPE}}$, Horseshoe again stands out with an average of 0.71\%, followed by NS (0.97\%), Ridge (1.06\%), Lasso (1.24\%), and Normal (1.37\%). This pattern suggests that Horseshoe effectively suppresses many small coefficients without overshrinking the larger signals, yielding robust relative‐error forecasts even as \(p\) grows. By contrast, Normal’s wider prior and Lasso’s strong \(\ell_1\) shrinkage can lead to higher MAPE in certain lag settings (see Figure~\ref{fig:canada_rmses_mapes_series}). 

\begin{figure}[ht!]
\centering
\includegraphics[width=1\textwidth]{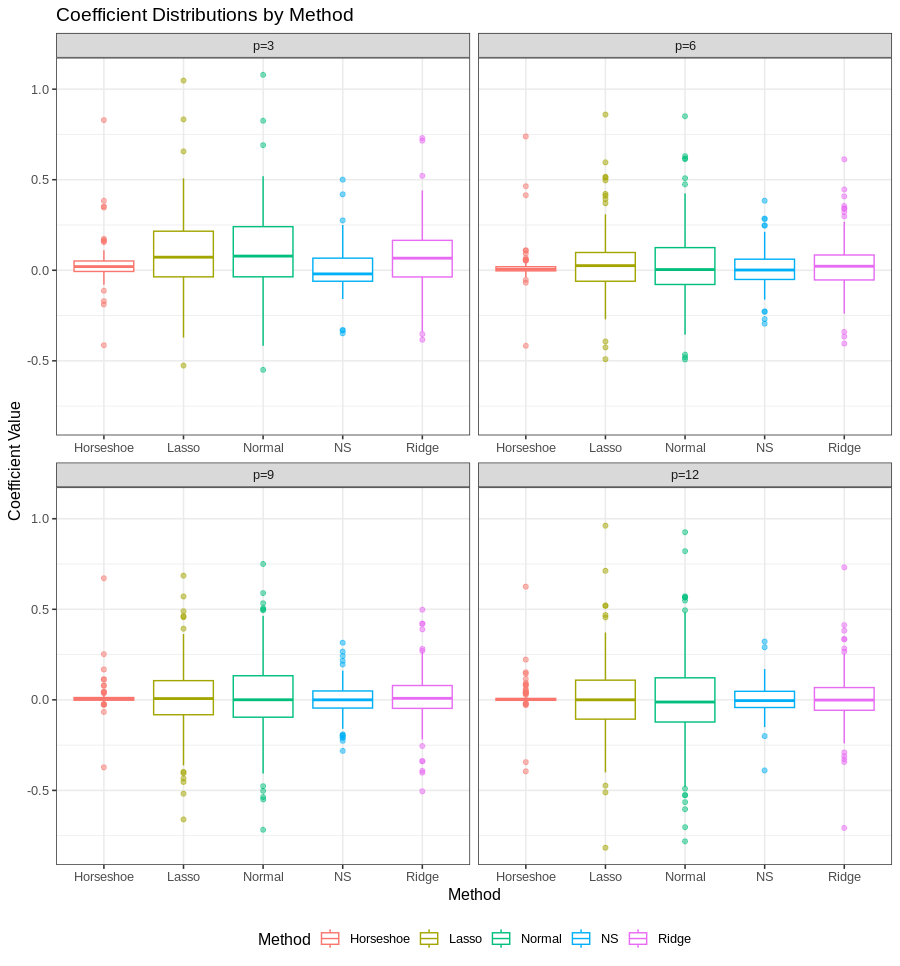}
\caption{Distribution of estimated VAR coefficients by method. Each point represents one of the $(4 \times 4 \times p)$ parameters, highlighting the degree of shrinkage for each prior.}
\label{fig:coefs_boxplot}
\end{figure}

Shrinkage patterns for a few representative lag orders (\(p=3,6,9,12\)) appear in Figure~\ref{fig:coefs_boxplot}. Horseshoe and Lasso consistently display heavy shrinkage toward zero for small coefficients, Normal has moderate Gaussian-like shrinkage, NS exhibits a broader coefficient spread, and Ridge pulls estimates closer to zero but never to an exact zero. Across the different $p$ values, Horseshoe’s local-global prior structure appears to adapt more flexibly, resulting in better overall forecasting metrics.

\subsubsection{Case Study: \texorpdfstring{$p=11$}{p=11}}
To illustrate performance at a higher lag, we examine the VAR(11) specification, which delivers the lowest average forecasting errors among the tested orders. Table~\ref{tab:results_table} shows each method’s RMSE and MAPE on the final four quarters of the holdout set. Notably, \emph{Horseshoe} again achieves the best performance on both measures (RMSE = 0.51, MAPE = 0.60\%). The next closest method is \emph{Ridge} (RMSE = 0.61, MAPE = 1.66\%) and \emph{NS} (RMSE = 0.65, MAPE = 1.26\%), while Lasso and Normal both exhibit slightly higher errors (RMSE = 0.70--0.78, MAPE = 1.66--1.81\%). This example highlights Horseshoe’s ability to preserve large coefficients and aggressively shrink small ones, maintaining strong predictive accuracy even at high lag orders.

\begin{figure}[ht!]
\centering
\includegraphics[width=1\textwidth]{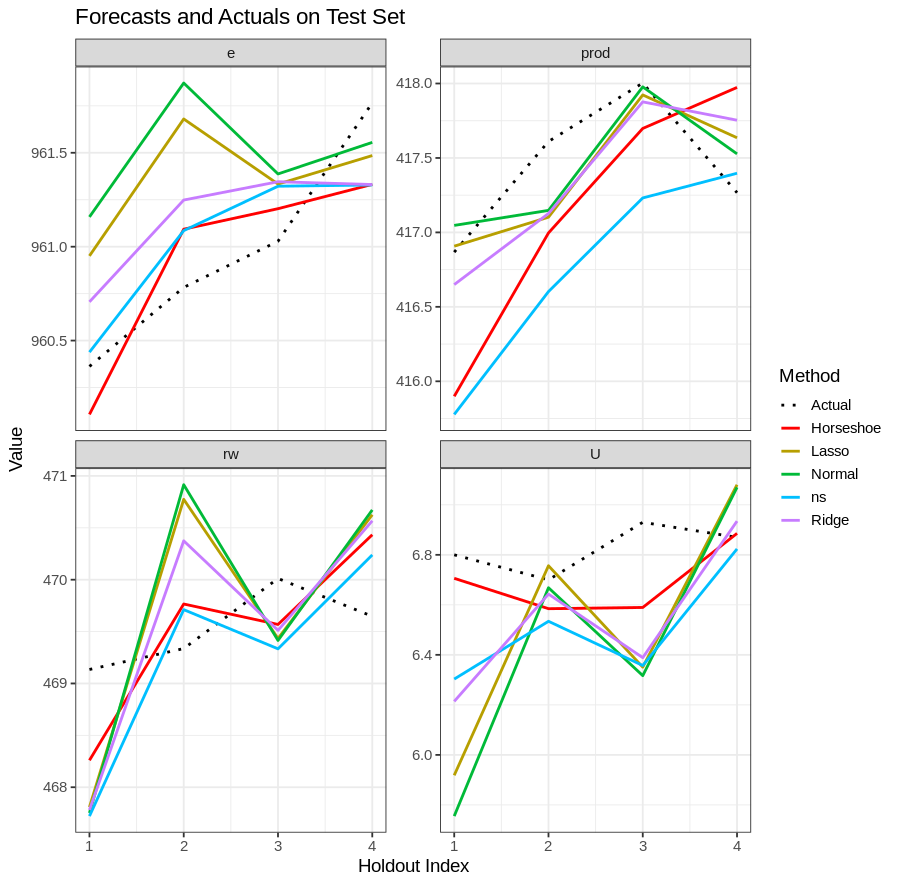}
\caption{Out-of-sample 1 step ahead forecasts from VAR(11) and actuals for each Canadian macroeconomic variable—employment ($e$), real wages ($rw$), productivity ($prod$), and unemployment ($U$).}
\label{fig:forecast_plot}
\end{figure}

 The 1-step-ahead forecasts of the observed data across the four holdout quarters are shown in figure~\ref{fig:forecast_plot}. Horseshoe, ns, and Ridge all track the actual values fairly closely. Lasso and Normal lag behind somewhat. Overall, these VAR($11$) results echo our broader simulation findings: Horseshoe’s adaptive local-global prior can maintain strong performance at high lag orders, and Ridge remains reasonably robust as well, whereas ns, Lasso, and Normal can become less accurate or more variable depending on the specific error metric.

\subsubsection{Practical Implications}

For \emph{unemployment}, the Horseshoe prior cuts the one‑step‑ahead RMSE from 0.63 to 0.51 (–19\%), reducing the average error at a 6 \% jobless rate from roughly 0.45 pp to 0.36 pp.  
System‑wide, it lowers the mean RMSE from 0.63 to 0.51 and roughly halves MAPE from 1.37 \% to 0.71 \% (Table \ref{tab:canada_summary_p1to10}), delivering noticeably sharper short‑term forecasts for all four Canadian macro‑series.

Relative to the strongest non‑Bayesian competitor, Ridge, Horseshoe still trims about 0.05 RMSE points and cuts MAPE by nearly one‑third.  Forecast variability across lag orders remains comparable (SD RMSE 0.22 for Horseshoe vs 0.24 for Ridge; Table \ref{tab:canada_summary_p1to10}), underscoring that the accuracy gains are achieved without sacrificing stability.

\section{Discussion}
\label{sec:discussion}

Our simulation results lead to several important takeaways about shrinkage estimation in VAR models under varying dimensionality and lag orders. First, the Horseshoe prior stands out for consistently achieving the lowest parameter RMSE and near-nominal coverage, particularly in the most challenging high-dimensional or overfit scenarios. This local-global prior structure successfully suppresses small coefficients while preserving truly large effects, thereby producing stable estimates and competitive forecasts across the board. By contrast, Lasso and Normal priors often deliver mid-range forecast accuracy and parameter estimation, but they maintain coverage near or above the 95\% target, albeit with wider intervals in some cases.

Ridge regression remains effective for forecasting in low- to moderate-dimensional scenarios (where the ratio of parameters to observations is not excessively large), frequently ranking second or third in terms of forecast RMSE. However, it underestimates parameter uncertainty in high-dimensional settings e.g., when the dimension-to-sample-size ratio is particularly large, leading to undercoverage. Similarly, Nonparametric Shrinkage (\textit{ns}) provides very short intervals and can achieve strong point forecasts, but it exhibits markedly low coverage in the same high-dimensional regimes, suggesting that its narrower intervals are overly optimistic about uncertainty in heavily over-parameterized models.

In the Canadian macro-economic data application, similar patterns emerge: Horseshoe and Ridge each exhibit strong one-step-ahead forecast accuracy, while Lasso, Normal, and \textit{ns} occasionally lag behind, particularly when the model order is large. Overall, these findings reinforce the benefits of using local-global shrinkage to adapt to large model spaces, especially for practitioners seeking reliable inference and coverage. Frequentist options like Ridge can still perform competitively in lower-dimensional or less overfit settings but risk severe undercoverage when the parameter space grows. 

Taken together, these results underscore that when parameter interpretation and interval validity are paramount, Horseshoe or other local-global Bayesian priors are well-suited to handle high-dimensional or inflated-lag VAR models. If short-term predictive performance alone is the principal goal, Ridge can remain attractive, provided one is willing to accept somewhat lower coverage in complex settings. The Lasso and Normal priors offer middle-ground alternatives, balancing coverage and moderate forecasting performance without fully matching Horseshoe’s combination of shrinkage strength and coverage reliability.

\subsection*{Data and Code Availability}
All \texttt{R} scripts and \texttt{Stan} model files used in this study are publicly available at \\
\href{https://github.com/harrisonekatz/BayesVAR-SimStudy}{https://github.com/harrisonekatz/BayesVAR-SimStudy}. 
In particular, the main simulation script 
\texttt{var\_three\_sim\_script.R} 
(which orchestrates data generation, frequentist and Bayesian estimation, and result collation) may be found in the repository’s \texttt{R/} directory. 
The repository also includes each of the \texttt{Stan} model files 
(\texttt{var\_normal.stan}, \texttt{var\_lasso.stan}, \texttt{var\_horseshoe.stan}), 
along with examples illustrating their usage. 
All results and figures in this manuscript can be reproduced by running the scripts found in that repository.

\subsection*{Conflict of Interest Disclosure} The authors declare that there are no conflicts of interest.

\section*{Figures and tables}

\begin{figure}[ht!]
\centering
\includegraphics[width=1\textwidth]{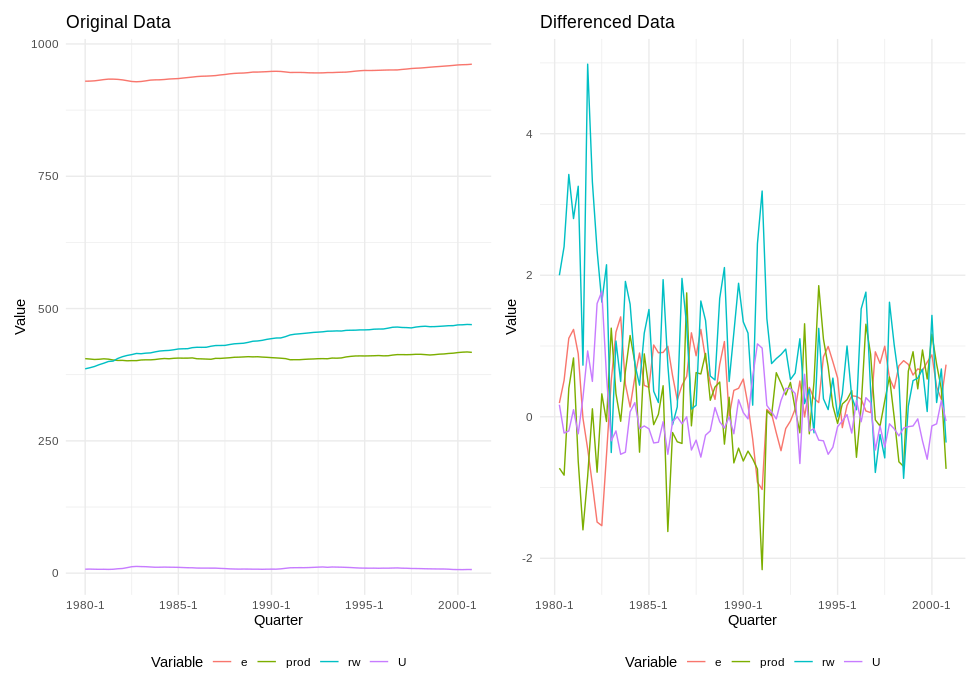}
\caption{%
The four Canadian macroeconomic variables in their original form (left) and once-differenced (right). Differencing helps remove trends and stabilize the series prior to VAR model estimation.
}
\label{fig:orig_vs_diff}
\end{figure}

\begin{figure}[ht!]
    \centering
    \includegraphics[width=1.1\textwidth]{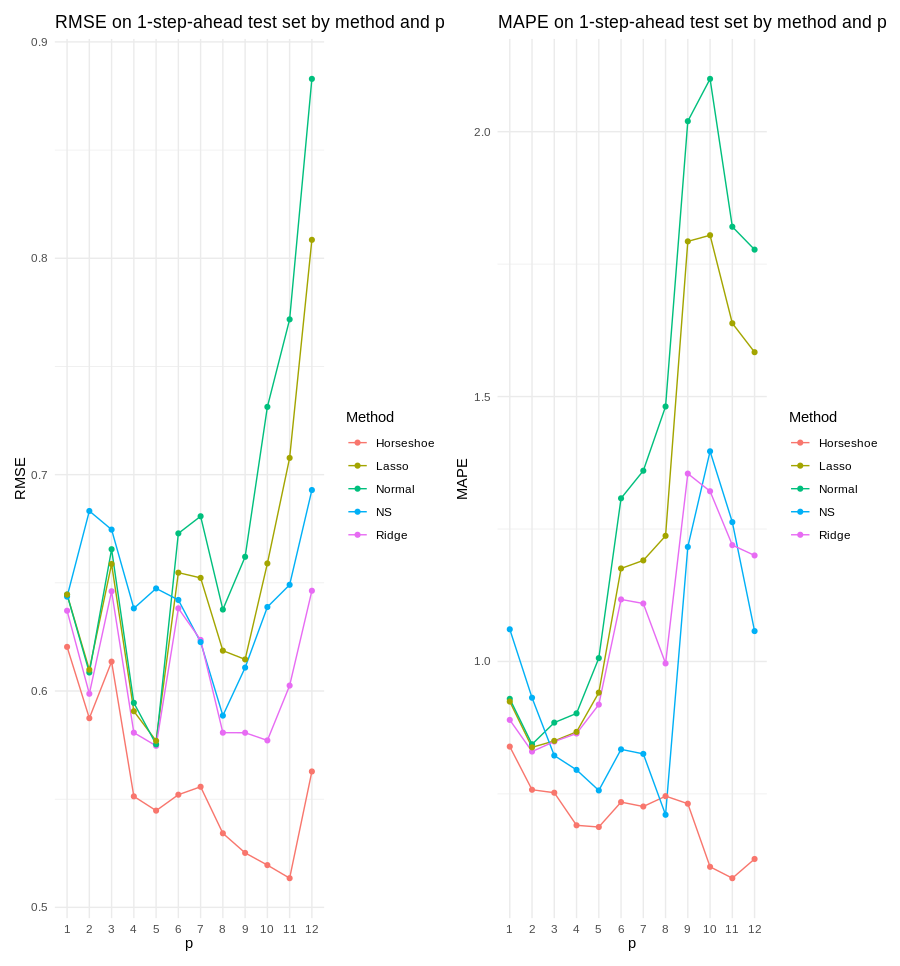}
    \caption{%
    \textbf{One‑step‑ahead forecast errors on the Canadian hold‑out sample (2000Q1--Q4) as the VAR lag order $p$ varies.} 
    \emph{Left:} root mean‑squared error $\mathrm{RMSE}_{p}$, averaged over the four macroeconomic series; 
    \emph{Right:} mean absolute percentage error $\mathrm{MAPE}_{p}$ (\%) averaged over the four macroeconomic series.  
    Lines trace the error obtained by each shrinkage estimator (Horseshoe, Ridge, Non‑parametric Shrinkage\,(NS), Lasso, Normal) for $p=1,\dots,12$.  
    Horseshoe delivers the lowest and most stable errors across all lag orders.  
    Ridge and NS perform competitively up to about $p\!\approx\!6$ but deteriorate more slowly thereafter, while Normal and Lasso show a pronounced rise in both RMSE and MAPE once $p\!>\!7$, reflecting over‑parameterization at high lag orders.}
    \label{fig:canada_rmses_mapes}
\end{figure}

\clearpage
\begin{table}[ht!]
\centering
\caption{%
\textbf{Overall Performance (All Coefficients).}
Mean and standard deviation (SD) of forecast RMSE (FRMSE), mean and SD of parameter RMSE (PRMSE), mean coverage (Cov), and mean interval length (Int.\ Length) across the three studies.
}
\label{tab:summary_all}
\begin{tabular}{llcccccc}
\toprule
\textbf{Scenario} 
& \textbf{Method}
& \textbf{FRMSE} & \textbf{FRMSE} 
& \textbf{PRMSE} & \textbf{PRMSE}
& \textbf{Cov}
& \textbf{Int.\ Length} \\
& 
& \textbf{Mean} & \textbf{SD}
& \textbf{Mean} & \textbf{SD}
& 
& \\
\midrule
\multicolumn{8}{c}{\emph{Study~1}} \\
\midrule
\multirow{5}{*}{Study~1}
 & Horseshoe & 0.211 & 0.0208 & 0.0434 & 0.0105 & 0.972 & 0.189 \\
 & Lasso     & 0.214 & 0.0217 & 0.0803 & 0.0111 & 0.943 & 0.311 \\
 & Normal    & 0.215 & 0.0218 & 0.0838 & 0.0115 & 0.936 & 0.318 \\
 & ns        & 0.213 & 0.0212 & 0.0693 & 0.00781 & 0.857 & 0.204 \\
 & Ridge     & 0.213 & 0.0215 & 0.0730 & 0.00897 & 0.904 & 0.261 \\
\midrule
\multicolumn{8}{c}{\emph{Study~2}} \\
\midrule
\multirow{5}{*}{Study~2}
 & Horseshoe & 0.325 & 0.0122 & 0.0536 & 0.00330 & 0.947 & 0.202 \\
 & Lasso     & 0.326 & 0.0124 & 0.0568 & 0.00319 & 0.951 & 0.223 \\
 & Normal    & 0.327 & 0.0125 & 0.0577 & 0.00326 & 0.949 & 0.225 \\
 & ns        & 0.327 & 0.0132 & 0.0598 & 0.00319 & 0.944 & 0.243 \\
 & Ridge     & 0.327 & 0.0129 & 0.0591 & 0.00487 & 0.840 & 0.173 \\
\midrule
\multicolumn{8}{c}{\emph{Study~3}} \\
\midrule
\multirow{5}{*}{Study~3}
 & Horseshoe & 0.342 & 0.0132 & 0.0394 & 0.00229 & 0.975 & 0.167 \\
  & Lasso     & 0.404 & 0.0194 & 0.104  & 0.00368 & 0.963 & 0.432 \\
 & Normal    & 0.418 & 0.0208 & 0.117  & 0.00446 & 0.955 & 0.464 \\
 & ns        & 0.365 & 0.0164 & 0.0619 & 0.00291 & 0.882 & 0.197 \\
 & Ridge     & 0.366 & 0.0153 & 0.0635 & 0.00206 & 0.845 & 0.181 \\
\bottomrule
\end{tabular}
\end{table}

\begin{table}[ht!]
\centering
\caption{%
\textbf{Performance on Zero Coefficients Only.}
Mean and SD of parameter RMSE (PRMSE) for zero coefficients, mean coverage (Cov), and mean interval length (Len).
}
\label{tab:summary_zero}
\begin{tabular}{llcccc}
\toprule
\textbf{Scenario} 
& \textbf{Method} 
& \textbf{PRMSE} & \textbf{PRMSE}
& \textbf{Cov} & \textbf{Len} \\
& 
& \textbf{Mean} & \textbf{SD}
& 
& \\
\midrule
\multicolumn{6}{c}{\emph{Study~1}} \\
\midrule
\multirow{5}{*}{Study~1}
 & Horseshoe & 0.0266 & 0.0082 & 0.999 & 0.176 \\
 & Lasso     & 0.0798 & 0.0119  & 0.945 & 0.313 \\
 & Normal    & 0.0839 & 0.0124  & 0.935 & 0.320 \\
 & ns        & 0.0606 & 0.0086 & 0.890 & 0.201 \\
 & Ridge     & 0.0711 & 0.0096 & 0.912 & 0.262 \\
\midrule
\multicolumn{6}{c}{\emph{Study~2}} \\
\midrule
\multirow{5}{*}{Study~2}
 & Horseshoe & 0.0357 & 0.0034 & 0.990 & 0.177 \\
 & Lasso     & 0.0558 & 0.0047 & 0.956 & 0.221 \\
 & Normal    & 0.0578 & 0.0049 & 0.950 & 0.225 \\
 & ns        & 0.0489 & 0.0039 & 0.978 & 0.240 \\
 & Ridge     & 0.0495 & 0.0037 & 0.904 & 0.173 \\
\midrule
\multicolumn{6}{c}{\emph{Study~3}} \\
\midrule
\multirow{5}{*}{Study~3}
 & Horseshoe & 0.020 & 0.002 & 1.000 & 0.152 \\
  & Lasso     & 0.107  & 0.004 & 0.965 & 0.445 \\
 & Normal    & 0.120  & 0.005 & 0.955 & 0.478 \\
 & ns        & 0.052 & 0.005 & 0.928 & 0.196 \\
 & Ridge     & 0.056 & 0.004 & 0.882 & 0.180 \\
\bottomrule
\end{tabular}
\end{table}
\begin{table}[ht!]
\centering
\caption{%
\textbf{Performance on Nonzero Coefficients Only.}
Mean and SD of parameter RMSE (PRMSE) for nonzero coefficients, mean coverage (Cov), and mean interval length (Len).
}
\label{tab:summary_nonzero}
\begin{tabular}{llcccc}
\toprule
\textbf{Scenario} 
& \textbf{Method} 
& \textbf{PRMSE} & \textbf{PRMSE}
& \textbf{Cov} & \textbf{Len} \\
& 
& \textbf{Mean} & \textbf{SD}
& 
& \\
\midrule
\multicolumn{6}{c}{\emph{Study~1}} \\
\midrule
\multirow{5}{*}{Study~1}
 & Horseshoe & 0.085 & 0.027 & 0.84 & 0.26 \\
 & Lasso     & 0.079 & 0.022 & 0.94 & 0.30 \\
 & Normal    & 0.080 & 0.023 & 0.94 & 0.31 \\
 & ns        & 0.098 & 0.028 & 0.70 & 0.22 \\
 & Ridge     & 0.078 & 0.023 & 0.86 & 0.25 \\
\midrule
\multicolumn{6}{c}{\emph{Study~2}} \\
\midrule
\multirow{5}{*}{Study~2}
 & Horseshoe & 0.0596 & 0.00391 & 0.929 & 0.213 \\
 & Lasso     & 0.0571 & 0.00327 & 0.949 & 0.224 \\
 & Normal    & 0.0576 & 0.00329 & 0.949 & 0.225 \\
 & ns        & 0.0639 & 0.00396 & 0.929 & 0.245 \\
 & Ridge     & 0.0626 & 0.00587 & 0.814 & 0.173 \\
\midrule
\multicolumn{6}{c}{\emph{Study~3}} \\
\midrule
\multirow{5}{*}{Study~3}
 & Horseshoe & 0.083 & 0.0049 & 0.86 & 0.24 \\
  & Lasso     & 0.092 & 0.0040 & 0.96 & 0.37 \\
 & Normal    & 0.098 & 0.0044 & 0.96 & 0.39 \\
 & ns        & 0.094 & 0.0063 & 0.67 & 0.20 \\
 & Ridge     & 0.088 & 0.0082 & 0.67 & 0.18 \\
\bottomrule
\end{tabular}
\end{table}

\begin{table}[ht!]
\centering
\caption{%
\textbf{Times Each Method Is ``Best'' in Forecast or Parameter RMSE.}
For each scenario and replication (50 total), we identify which method attains the lowest forecast RMSE or lowest parameter RMSE. 
Columns show the percentage of replications in which each method is best.
}
\label{tab:times_best}
\begin{tabular}{l l c}
\toprule
\multicolumn{3}{c}{\textbf{Best in Forecast RMSE}} \\
\midrule
\textbf{Scenario} & \textbf{Method} & \textbf{\% of Replications} \\
\midrule
\multirow{5}{*}{Study~1} 
 & Horseshoe & $60\%$ \\
 & ns       & $20\%$ \\
 & Normal   & $10\%$ \\
 & Ridge    & $6\%$ \\
 & Lasso    & $4\%$ \\
\midrule
\multirow{5}{*}{Study~2}
 & Horseshoe & $48\%$ \\
 & Lasso    & $20\%$ \\
 & ns       & $20\%$ \\
 & Ridge    & $10\%$ \\
 & Normal   &  $2\%$ \\
\midrule
\multirow{1}{*}{Study~3} 
 & Horseshoe & $100\%$ \\
\bottomrule
\\[6pt]
\toprule
\multicolumn{3}{c}{\textbf{Best in Parameter RMSE (All Coefficients)}} \\
\midrule
\textbf{Scenario} & \textbf{Method} & \textbf{\% of Replications} \\
\midrule
Study~1 & Horseshoe & $100\%$ \\
\hline
\multirow{2}{*}{Study~2} 
  & Horseshoe & $90\%$ \\
  & Ridge     & $10\%$ \\
  \hline
Study~3 & Horseshoe & $100\%$ \\
\bottomrule
\end{tabular}
\end{table}

\begin{table}[ht!]
\centering
\caption{%
\textbf{Forecast error summaries for VAR($p$), $p=1,\dots,12$.}
Shown are the mean and standard deviation of RMSE and MAPE (\%) across the 12 lag choices. Horseshoe achieves the smallest mean RMSE and MAPE, while Normal exhibits the largest mean RMSE and MAPE. Ridge, NS, and Lasso provide intermediate performance.}
\begin{tabular}{lcccc}
\toprule
\textbf{Method} &
$\boldsymbol{\overline{\mathrm{RMSE}}}$ &
\textbf{SD\,RMSE} &
$\boldsymbol{\overline{\mathrm{MAPE}}}$ &
\textbf{SD\,MAPE} \\
\midrule
Horseshoe & 0.51 & 0.22 & 0.71 & 0.94 \\
Lasso     & 0.60 & 0.25 & 1.24  & 1.87  \\
NS        & 0.60 & 0.25 & 0.97 & 1.39  \\
Normal    & 0.63 & 0.26 & 1.37  & 2.12  \\
Ridge     & 0.56 & 0.24 & 1.06  & 1.51  \\
\bottomrule
\end{tabular}

\label{tab:canada_summary_p1to10}
\end{table}

\begin{table}[ht!]
\centering
\caption{Forecasting accuracy on the Canada data for the VAR(11) model, evaluated on the final four observations. Lower RMSE and MAPE values indicate better performance.}
\begin{tabular}{lcc}
\hline
\textbf{Method} & \textbf{RMSE} & \textbf{MAPE (\%)} \\
\hline
Horseshoe       & 0.51          & 0.60 \\
Lasso           &  0.70        & 1.66 \\
Normal          &  0.78        & 1.81 \\
ns                & 0.65          & 1.26 \\
Ridge           & 0.61        & 1.66 \\

\hline
\end{tabular}

\label{tab:results_table}
\end{table}

\clearpage

\bibliographystyle{chicago}
\bibliography{references}

\end{document}